\documentclass[aps,prl, superscriptaddress, preprint, floatfix]{revtex4-1}

\usepackage{here}
\usepackage{graphicx}
\usepackage{wrapfig}
\usepackage{sidecap}

\usepackage{multirow}

\usepackage{amsmath}
\usepackage{amssymb}
\usepackage{braket}

\usepackage{array}

\usepackage{hyperref}

\begin{document}

\title{Nonlinear electron-phonon coupling in doped manganites}
\date{\today}

\author{V. Esposito}
\email[Corresponding author: ]{vincent.esposito@psi.ch}
\affiliation{Swiss Light Source, Paul Scherrer Institut, 5232 Villigen PSI, Switzerland}

\author{R. Mankowsky}
\affiliation{Max-Planck Institute for the Structure and Dynamics of Matter, 22761 Hamburg, Germany}
\affiliation{Center for Free Electron Laser Science, 22761 Hamburg, Germany}

\author{M. Fechner}
\affiliation{Materials Theory, ETH Z\"{u}rich, Wolfgang-Pauli-Strasse 27, 8093 Z\"{u}rich, Switzerland}

\author{H. Lemke}
\affiliation{LCLS, SLAC National Accelerator Laboratory, Menlo Park, California 94025, USA}
\affiliation{SwissFEL, Paul Scherrer Institut, 5232 Villigen PSI, Switzerland}

\author{M. Chollet}
\affiliation{LCLS, SLAC National Accelerator Laboratory, Menlo Park, California 94025, USA}
\author{J.M. Glownia}
\affiliation{LCLS, SLAC National Accelerator Laboratory, Menlo Park, California 94025, USA}

\author{M. Nakamura}
\affiliation{RIKEN Center for Emergent Matter Science, Wako 351-0198, Japan}

\author{M. Kawasaki}
\affiliation{RIKEN Center for Emergent Matter Science, Wako 351-0198, Japan}
\affiliation{Department of Applied Physics and Quantum Phase Electronics Center (QPEC), University of Tokyo, Tokyo 113-8656, Japan}

\author{Y. Tokura}
\affiliation{RIKEN Center for Emergent Matter Science, Wako 351-0198, Japan}
\affiliation{Department of Applied Physics and Quantum Phase Electronics Center (QPEC), University of Tokyo, Tokyo 113-8656, Japan}

\author{U. Staub}
\affiliation{Swiss Light Source, Paul Scherrer Institut, 5232 Villigen PSI, Switzerland}

\author{P. Beaud}
\affiliation{Swiss Light Source, Paul Scherrer Institut, 5232 Villigen PSI, Switzerland}
\affiliation{SwissFEL, Paul Scherrer Institut, 5232 Villigen PSI, Switzerland}

\author{M. F\"{o}rst}
\affiliation{Max-Planck Institute for the Structure and Dynamics of Matter, 22761 Hamburg, Germany}
\affiliation{Center for Free Electron Laser Science, 22761 Hamburg, Germany}

\begin{abstract}
We employ time-resolved resonant x-ray diffraction to study the melting of charge order and the associated insulator-metal transition in the doped manganite Pr$_{0.5}$Ca$_{0.5}$MnO$_{3}$ after resonant excitation of a high-frequency infrared-active lattice mode. We find that the charge order reduces promptly and highly nonlinearly as function of excitation fluence. Density functional theory calculations suggest that direct anharmonic coupling between the excited lattice mode and the electronic structure drive these dynamics, highlighting a new avenue of nonlinear phonon control.
\end{abstract}

\maketitle

Some of the most fascinating phenomena in condensed matter physics arise from electron-phonon interactions. A striking example is the BCS theory for superconductivity, where phonons mediate an attractive interaction between two electrons with opposite spin and momentum to allow the carriers to condensate into the superconducting state \cite{Bardeen1957}. In metals and semiconductors, transport properties are also shaped by linear electron-phonon (e-ph) coupling, in particular through the creation of polarons \cite{Mahan2000}. Furthermore, e-ph coupling plays an important role in the physics of perovskite oxides such as the mixed-valence manganites, where a variety of electronic and magnetic phases is stabilized via the Jahn-Teller effect \cite{Dagotto2003}. The precarious equilibrium between these possible ground states is easily perturbed by external stimuli such as static electric or magnetic fields, temperature, pressure or even by light, opening new ways of manipulating matter on ever faster timescales with short laser pulses \cite{Miyano1997, Schmitt2008, Kampfrath2011, Coslovich2013, Mankowsky2014, Kubacka2014, Beaud2014}.

The recent progress in the generation of high-energy ultrashort pulses in the mid-infrared (mid-IR) range permits to resonantly excite vibrational modes of the lattice to large amplitudes, exceeding several percent of the interatomic distances --- a value at which nonlinear coupling of phonons to other degrees of freedom becomes important \cite{Rini2007, Fausti2011, Forst2011}. This approach enabled mode-selective material control on sub-picosecond timescales, as exemplified in the possible enhancement of superconductivity \cite{Mankowsky2014, Fausti2011}, the occurrence of insulator-metal transitions and the suppression of magnetic and orbital order in manganites \cite{Rini2007, Forst2011a}.
A proposed mechanism, referred to as 'nonlinear phononics' and based on nonlinear phonon-phonon coupling, suggests that these phenomena result from the rectification of the excited mode and a net displacement of the crystal lattice along the coordinates of anharmonically coupled vibrational modes, which control the electronic properties \cite{Mankowsky2014,Forst2011a, Subedi2014, Fechner2016}. However, a possible direct coupling between the excited mode and the electronic system has only seldomly been considered \cite{Kim2016}, although, for example, the vibrational excitation of a molecular solid was shown to coherently perturb the electronic interactions by directly affecting the orbital wave functions \cite{Singla2015}.

In this Letter, we explore the nonlinear electron-phonon coupling in a manganite by investigating the lattice driven ultrafast insulator-metal transition. We use resonant x-ray diffraction at the Mn \textit{K} edge to study the dynamics of the electronic and structural order in a Pr$_{0.5}$Ca$_{0.5}$MnO$_{3}$ (PCMO) thin film induced by the large-amplitude excitation of a vibrational mode with 200 fs mid-IR pulses. We then compare our experimental findings with \textit{ab initio} calculations of the relevant couplings.

\begin{figure*}
\centering
\includegraphics[scale=1]{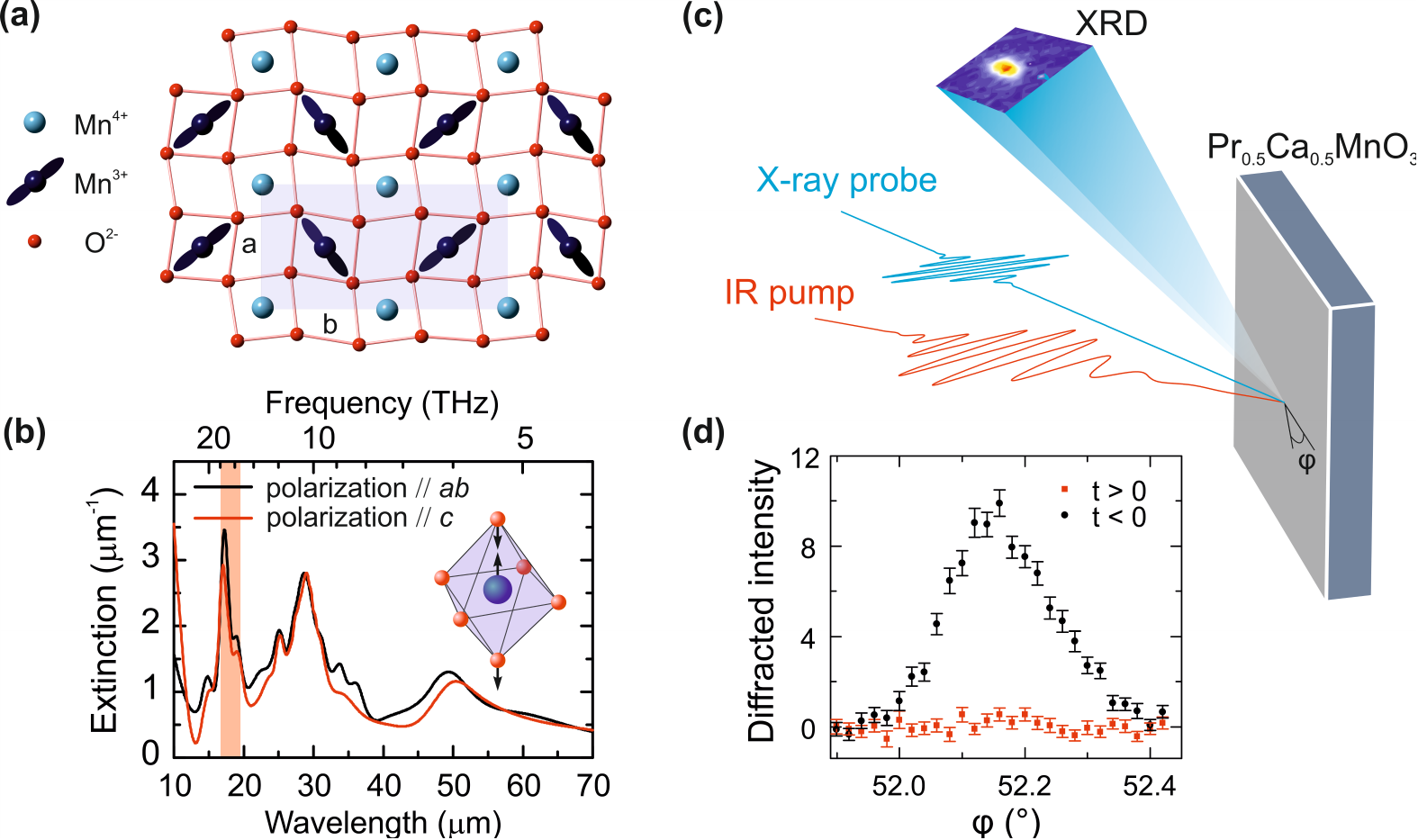}
\caption{(a) Charge and orbital order in PCMO in the \textit{ab}-plane. The low temperature unit cell is delimited by the blue shaded area.  (b) Absorption spectra of a Nd$_{0.5}$Ca$_{0.5}$MnO$_{3}$ single crystal in the mid-IR range (reproduced from ref. \cite{Tobe2004}). This materials displays a very similar phase diagram than PCMO and has the same structure and COO pattern below T$_{CO}$. The shaded red area represents spectral range of the pump. The inset shows the dominant motion driven by the excitation. (c) Experimental setup at the XPP beamline at the LCLS. The angle $\phi$ indicates the rotation angle around the sample's surface normal.  (d) Rotational scan around the sample surface normal (angle φ in panel c) of the $(0\bar{3}0)$ reflection before and 4 ps after excitation at a pump fluence of 10.7 mJ/cm$^2$. The complete suppression of the Bragg peak after photo-excitation demonstrates the melting of the COO.}
\label{Fig1}
\end{figure*}

At room temperature, Pr$_{0.5}$Ca$_{0.5}$MnO$_{3}$ is a paramagnetic semiconductor that undergoes a large distortion of the cubic perovskite structure. Charge and orbital order (COO) and antiferromagnetic (AFM) phases arise upon cooling below T$_{CO}$ = 240 K and T$_{N}$ = 150 K respectively \cite{Dagotto2003}. This so-called CE-type COO is characterized by a zig-zag arrangement of the \textit{3d e$_g$} orbitals and a checkerboard pattern of Mn$^{3+}$/Mn$^{4+}$ions (Fig. \ref{Fig1} (a)). The long range ordering of the electrons goes in hand with a structural distortion due to the Jahn-Teller effect at the Mn$^{3+}$ sites that lowers the crystal symmetry from orthorhombic \textit{Pbnm} to monoclinic $P2_1/m$ and a doubling of the unit cell along the \textit{b} axis, leading to weak superlattice reflections of the type $(h \frac{k}{2} l)$.

The 40 nm PCMO thin film was grown by pulsed laser deposition on a $(011)$-oriented LSAT substrate, as described in \cite{Okuyama2009} and has an ordering temperature T$_{CO}$ of 220 K, only slightly below the reported bulk critical temperature. During the experiment, the sample was kept at 100 K, well below T$_{CO}$ by means of a nitrogen cryo-blower. The sample was excited with 200 fs mid-IR pulses produced by mixing the signal and idler from a high energy optical parametric amplifier (OPA) seeded by the 800nm output of an amplified Ti:S laser. The OPA output wavelength was tuned to the stretching mode of the apical Mn-O bond ($\lambda \approx 17$ $\mu$m, bandwidth ≈ 1.5 $\mu$m) with a maximum energy per pulse of 33 $\mu$J (Fig. \ref{Fig1} (b)).
The subsequent dynamics were probed by resonant X-ray diffraction at the Mn K edge with monochromatized 50 fs pulses provided by the LCLS free electron laser at the SLAC National Accelerator Laboratory. To guarantee an optimum time resolution the arrival time difference between the pump and the probe was measured shot-to-shot by means of the spectral encoding technique \cite{Harmand2013}. A sketch of the pump-probe experiment that was carried out at the XPP instrument \cite{Chollet2015} is shown in Fig. \ref{Fig1} (c).
The probed area is homogenously excited due to the relatively large 280 x 750 $\mu$m$^2$ pump spot (versus 50 x 50 $\mu$m$^2$ for the X-ray beam). P-polarized light ensures maximum power transmitted into the sample. Because the penetration depth of the mid-IR (360 nm) is much larger than the film thickness (40 nm), the excitation density can be considered as uniform over the entire probed volume. Optimization of various geometrical constraints leads to the non-collinear geometry of the experiment (see Fig. \ref{Fig1} (c)), more details can be found in the Supplementary Informations.

The charge order response is measured by the intensity of the $(0\bar{3}0)$ reflection at the Mn \textit{K} edge, which directly relates to the charge disproportionation at the Mn sites \cite{Zimmermann1999}. Figure \ref{Fig1} (d) shows the disappearance of the $(0\bar{3}0)$ Bragg peak 200 femtoseconds after excitation at a fluence of 10.7 mJ/cm$^2$, demonstrating complete charge and orbital order melting.
The dynamics of the charge order peak for different excitation fluence are shown in Figure 2 (a). Minor intensity changes are observed below 5 mJ/cm$^2$. Above a critical fluence of f$_c = 8.7$ mJ/cm$^2$, the $(0\bar{3}0)$ reflection disappears within the experimental time resolution of ~200 fs, clearly evidencing prompt CO melting. In a narrow range of intermediate fluences, the CO is only partially melted and recovers over several tens of picoseconds. The fluence dependence of the intensity drop, shown in Figure 2 (b), highlights this nonlinear, threshold-like behavior. In contrast, above-bandgap-excitation at near-IR 800 nm wavelengths, which directly perturbs the electronic system, results in prompt charge order melting with a linear fluence dependence \cite{Beaud2014}, as shown in the overlaid data.

\begin{figure}
\centering
\includegraphics[scale=1]{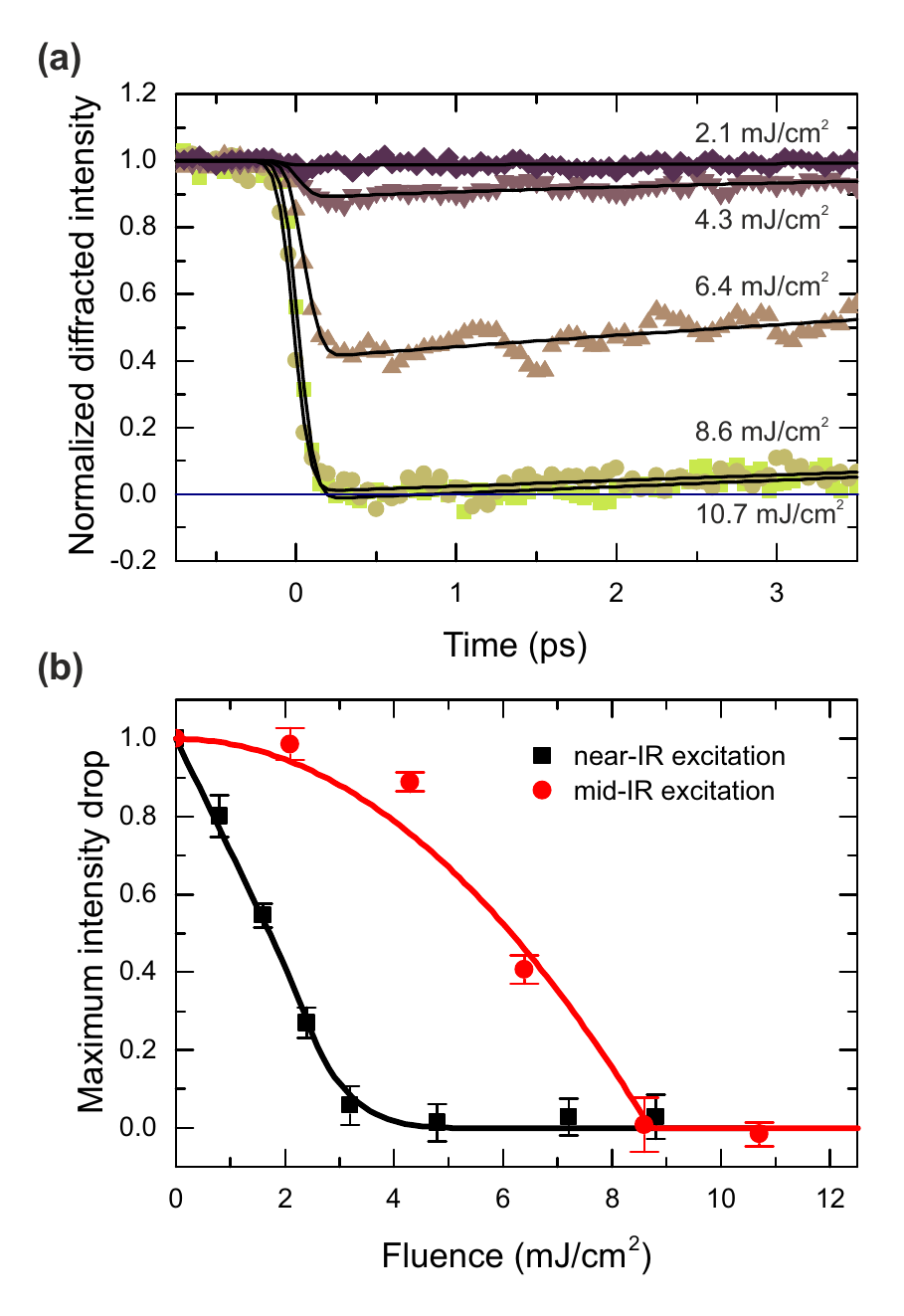}
\caption{Dynamic of charge order melting. (a) Transient response of the $(0\bar{3}0)$ charge order reflection for different fluence ($\lambda = 17$ $\mu$m). The time traces were taken at the maximum intensity of the Bragg peak (Fig. \ref{Fig1} (d) ). The error bars are about 1.5 times the symbol size and are not displayed for clarity. The black lines are fits to the data consisting of an error function with an exponential recovery. (b) Whereas excitation with near-IR pulses leads to a linear drop in diffraction intensity as a function of fluence and can be described by the Landau-type order parameter proposed in Ref. \cite{Beaud2014} (black squares and line), the intensity drop following mid-IR excitation is highly non-linear. The red circles are the maximal drop given from the fit of the time evolution traces and the red line is a fit using $1-\left( \frac{f}{f_c} \right)^2$  yielding a critical fluence of $8.7 \pm 0.7$ mJ/cm$^2$.}
\label{Fig2}
\end{figure}

We performed DFT calculations to identify the lattice dynamics induced by the intense mid-infrared excitation and their effects on the electronic structure. 
The first principle calculations are performed in the framework of density functional theory using the generalized gradient approximation (GGA) \cite{Perdew1996} as implemented within the Vienna ab-initio simulation package (VASP) \cite{Kresse1996}. All our calculations are done with the default projector augmented wave (PAW) pseudopotentials \cite{Blochl1994} which exhibit the following electronic configurations: Pr (6p$^{1}$, 6s$^{2}$, 5p$^{6}$,5s$^{2}$), Ca (3s$^{2}$,3p$^{6}$), Mn (3d$^{5}$, 4s$^{2}$, 3p$^{6}$) and O (2s$^{2}$, 2p$^{4}$). To take into account strong correlation effects, we employ a \textit{DFT+U} scheme \cite{Anisimov1997} with U=5.0 eV and J=0.0 eV applied to the Mn d-states. We select this value to reproduce the experimental band gap of 0.37 eV \cite{Okimoto1999}. After testing the convergence of forces, phonon frequencies and anharmonic coupling constants, we chose a 3x5x7 \textit{k}-point mesh in combination with a cutoff energy for the plane wave basis set of 550 eV. We further perform structural relaxations, to obtain force free reference structures, required for computations of the phonon spectra. The convergence thresholds for the electronic and ionic steps are set to 10$^{-8}$ eV and 0.1 meV/\r{A} respectively. More details on the phonon calculations are given in the Supplementary Informations.
The analysis of the zone-center phonon modes reveals that only two infrared-active (IR) modes at 17.2 and 18.9 THz are significantly excited by the applied mid-IR pulse. First, we consider the anharmonic coupling of these modes to Raman modes within the framework of ‘nonlinear phononics’. As opposed to Ref. 17, however, we use the low temperature charge ordered $P2_1/m$ structure (see Supplementary Informations for more details). Mapping the total energies of our DFT calculations onto the full phonon-phonon potential $V=\omega_{IR}^2 Q_{IR}^2+\omega_R^2 Q_R^2+a_3 Q_R Q_{IR}^2+...$, we find that both excited IR modes couple only weakly (as $a_3/\omega_R^2$) to Raman modes, displacing the lattice along the coordinates of low frequency Ag modes by maximum 0.05 $\sqrt{\mu}$\r{A} (see Supplementary Informations). This atomic motion is by more than one order of magnitudes smaller than observed in YBa$_2$Cu$_3$O$_{6.5}$, where the enhancement of superconductivity was shown to be the consequence of such lattice anharmonicity \cite{Mankowsky2014, Fechner2016}. Furthermore, the PCMO bandgap reduction induced by the atomic motions of the nonlinearly coupled A$_g$ Raman modes scales linearly with the excitation fluence (see Supplementary Informations), overall suggesting that the nonlinear phononics mechanism is likely not driving the insulator-metal transition.

\begin{figure}
\centering
\includegraphics[scale=1]{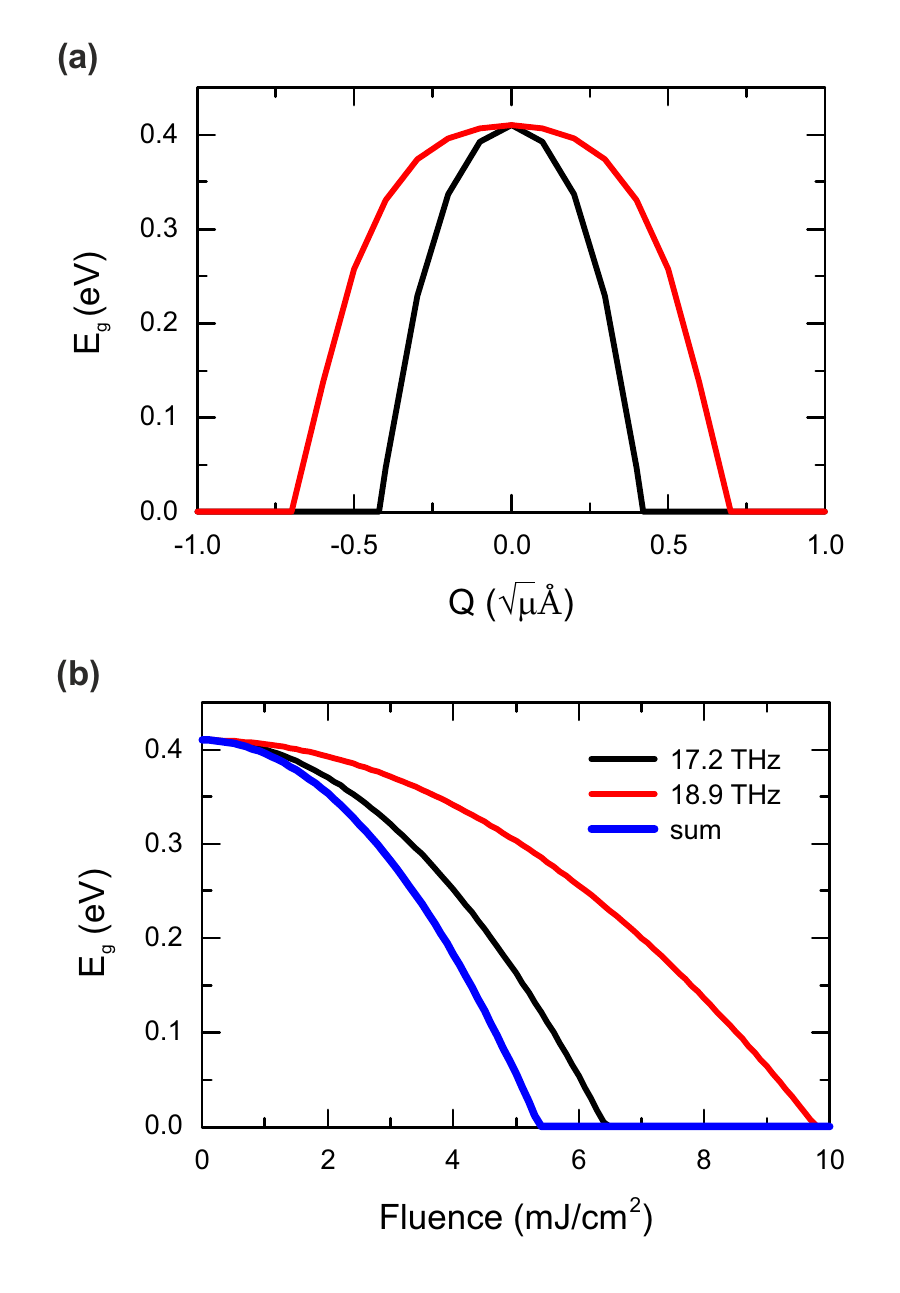}
\caption{(a) Calculated gap energy as a function of the frozen amplitude of the two excited IR modes. Interestingly, the quadratic coefficient also vanishes and the coupling is found to be purely quartic. The phonon amplitude is given in units of√μ A ̇, where μ is the reduced phonon mass. (b) The electronic gap as a function of mid-IR fluence for the two excited IR modes and their summed contribution. All other contributions, from additional IR modes or coupled Raman phonon are found to be negligible (see Suppl. Materials).}
\label{Fig3}
\end{figure}

Hence, we explore the changes of the electronic structure due to the atomic displacement along the coordinates of the directly excited IR mode. The CO state is stabilized by the creation of an electronic gap (E$_g$), whose lattice-induced alterations are computed within the frozen phonon approximation for both modes. We find a quartic reduction of the gap size as a function of both mode amplitudes, which at large values close the band gap as shown in Fig. \ref{Fig3} (a). 

To confront this finding with the experimental observation, we calculate the fluence dependent amplitudes of the IR modes by numerically solving the equations of motion \cite{Juraschek2016}. For simplicity, we assume that the electronic gap is determined by the maximum phonon amplitude $Q_{\text{IR,max}}^4$ of the directly driven phonon modes, which is reached at the end of the mid-IR excitation pulse. However, we cannot rule out that dynamically the gap size reduces with the time-average $ \langle Q_{\text{IR}}^4 \rangle$ of these modes.
Figure \ref{Fig3} (b) shows the calculated gap size for both individually excited modes and their summed contribution, revealing a critical pump fluence of $f_c = 5.4$ mJ/cm$^2$ for a full gap closure. This value is slightly lower than found experimentally. However, when taking into account the $32^{\circ}$ angle between the mid-IR pump electric field and the c-axis, the estimated experimental critical fluence reduces by 28\% to f$_c \approx ≈ 6.3$ mJ/cm$^2$, which is in excellent agreement with the calculation. As both the calculated gap energy and the measured (03 ̅0) Bragg intensity represent the square of the charge order parameter \cite{Ross2014}, these quantities can be directly compared. A quadratic function fit, as expected from the square reduction of the gap energy with excitation fluence, reproduces the measured diffracted intensity drop reasonably well (see Fig. \ref{Fig2} (b)).

\begin{figure}
\centering
\includegraphics[scale=1]{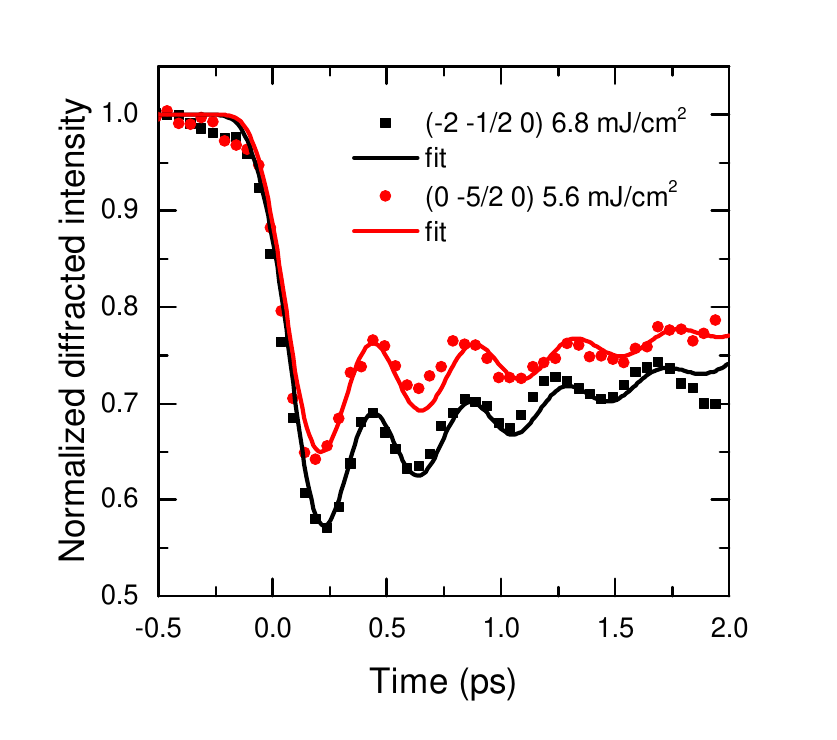}
\caption{Response of the structural superlattice reflection $(\bar{2} \bar{\frac{3}{2}}0)$ and  the orbital order reflection $(0\bar{\frac{5}{2}}0)$ at intermediate pump fluence. As for excitation of the electronic system at near-IR wavelength, a 2.4 THz coherent oscillation around a new equilibrium position is observed, suggesting a displacive excitation mechanism triggered by the fast melting of charge and orbital order in both cases. The solid lines are fits of an exponential recovery and a damped oscillation.}
\label{Fig4}
\end{figure}

We further probed the dynamics of the crystal lattice by measuring two superlattice reflections (Fig. \ref{Fig4}): the $(0\bar{\frac{5}{2}}0)$ peak, sensitive to the Jahn-Teller distortion and orbital order at the Mn$^{3+}$ sites, and the $(\bar{2} \bar{\frac{3}{2}}0)$, a direct measure of the overall structural distortion that accompanies the charge and orbital ordering. The fast drop in diffracted intensity and the coherent 2.4 THz modulations for both reflections indicate a displacive excitation of atomic motions towards the high-symmetry phase triggered by the melting of the CO/OO \cite{Beaud2014, Caviezel2013}. This structural relaxation closely resembles the dynamics following direct excitation of the electronic system, suggesting that in both cases the charge order melting precedes (and therefore initiates) the structural phase transition. Hence, one may conclude that mainly the Mott physics – as opposed to the Jahn-Teller effect – stabilize the ground state of this material.

The phenomenological model, describing charge order dynamics induced by exciting the electronic system in the near-IR, proposes the charge order parameter $\eta = \sqrt{1-\frac{n}{n_c} }$ in the driven state to depend solely on the excess energy in the electronic system \cite{Beaud2014}. A critical excitation density of 500 J/cm$^3$ was required for the phase transition to occur. Assuming linear extinction, direct lattice excitation investigated here requires a significantly lower critical excitation density of 200 J/cm$^3$ indicating, possibly, a more concerted way to reach the transient metallic state. However, once the gap becomes smaller than the photon energy, direct photo-doping into the electronic system will set in. The details of these dynamics could be experimentally investigated by monitoring the CO during the pump pulse, an experiment that requires carrier-envelope phase-stable mid-IR pulses and a high time resolution. Further insight on the energy balance between the lattice and the electrons will be gained once time-resolved dynamical calculations will become feasible for such large systems. In addition, a comprehensive characterization of the photo-induced transient state is still missing and other crucial aspects of the transition need to be explored. The spin dynamics and the magnetic ordering have been completely ignored here \cite{Li2013}, yet in the manganites transport properties are tightly bound to the magnetic state.

In summary, we have shown that the melting of the charge and orbital order and the associated metal-insulator transition following coherent lattice excitation in PCMO are driven by the direct and highly nonlinear coupling between the excited IR modes and the electronic degrees of freedom. As the gap closes, the charge and orbital order is suppressed and the Jahn-Teller distortion released, triggering structural relaxation. When combined with DFT calculations, the observed nonlinear behavior is well explained by the quartic dependence of the electronic gap on the excited phonon coordinate. This is an example of a new and direct way of phonon control of materials, where the electron-phonon interaction is pushed to the nonlinear regime, allowing the coupling with odd-parity modes.

This work was supported by the NCCR Molecular Ultrafast Science and Technology (NCCR MUST), research instrument of the Swiss National Science Foundation (SNSF). Use of the Linac Coherent Light Source (LCLS), SLAC National Accelerator Laboratory, is supported by the U.S. Department of Energy, Office of Science, Office of Basic Energy Sciences under Contract No. DE-AC02-76SF00515. M.N. was supported by the Japan Science and Technology Agency (JST), PRESTO.

\newpage

\vspace*{5cm}
\begin{center}
\textbf{{\Huge SUPPLEMENTARY INFORMATIONS}}
\end{center}
\newpage

\section*{Pump polarization and crystal orientation}
In order to excite a desired IR active phonon mode, the driving electric field has to point along a specific crystallographic direction. This is, however, not the only geometrical experimental constraint, as the crystal orientation in the laboratory frame is also defined by the diffraction condition for the reflection that is being probed. Because the pump polarization cannot be rotated (it is set horizontally in the laboratory frame) both criteria have to be fulfilled together. The only variable degrees of freedom left are then the incident angles of both beams. But other constraints are the time resolution which is optimized with collinear pump and probe beams and the diffracted intensity that is maximized at smaller grazing incidence angles. Although there is no general solution we can optimize the electric field of the pump light along a specific axis by carefully choosing the appropriate sample domain and incident angles. The pump polarization angle to the c axis is given in Table S1 for two possible domains with a mid-IR incident angle of 45$^{\circ}$, a grazing X-ray angle of 15$^{\circ}$ and at the $(0\bar{3}0)$ diffraction condition. This choice of incident angles only slightly increases the time resolution from 200 to 220 fs, while keeping a decent diffracted signal and the electric pump field polarized mainly along the desired c- or to the a-axis when looking at different domains in the film.

\begin{table}[H]
\centering
\begin{tabular}{p{1.8cm} p{1.8cm} p{1.8cm} p{1cm}}
\hline \hline
Domain & hkl & $\gamma$ & $\alpha$ \\ \hline
$(\bar{1} \bar{1} 2)$ & $(0\bar{3}0)$ & $32^{\circ}$ & $61^{\circ}$ \\
$(1 \bar{1} 2)$ & $(0\bar{3}0)$ & $82^{\circ}$ & $21^{\circ}$ \\
\hline \hline
\end{tabular}
\caption{Pump polarization angle to the axis of the crystal. The domain is defined by its surface normal; the angle $\gamma$ is the angle to the \textit{c}-axis and $\alpha$ is the angle to the \textit{a}-axis.}
\label{T1}
\end{table}

\section*{Pump polarization dependence of the photo-induced phase transition}
\begin{figure}
\centering
\includegraphics[scale=1.2]{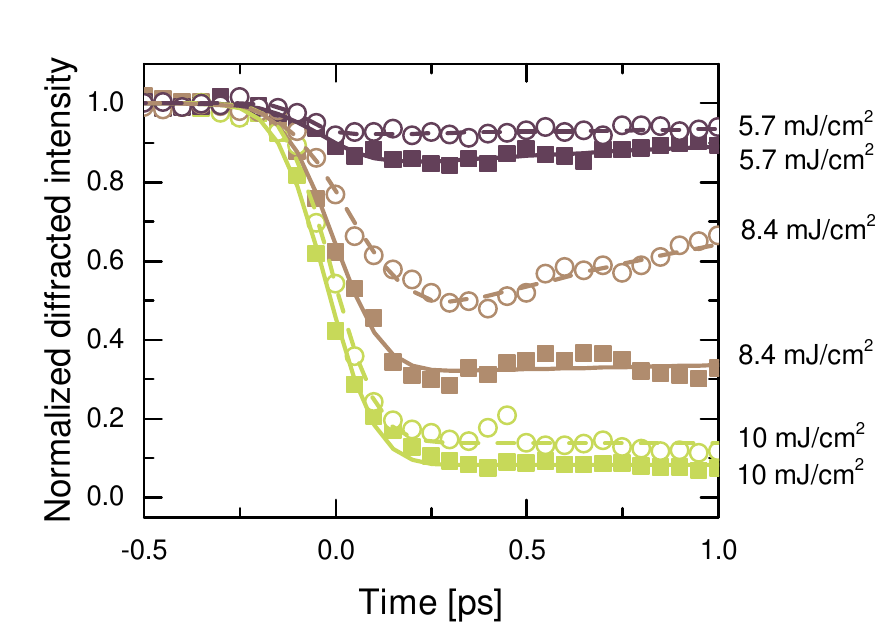}
\caption{Transient response of the $(0\bar{3}0)$ reflection upon excitation of orthogonal modes at 140 K. The full square symbols are for an excitation along the c-axis and the empty circles for an excitation in the ab-plane.}
\label{S1}
\end{figure}

While most of the experiment has been performed on the $(0\bar{3}0)$ peak of the $(\bar{1} \bar{1} 2)$ domain, data were also taken on the $(1 \bar{1} 2)$ domain, where the electric field of the pump is polarized in the \textit{ab}-plane (Fig. \ref{S1}). We observe a complete melting of the charge order for both directions of excitation. The excitation along the c-axis is, however, slightly more efficient than in the ab-plane, even more so when considering that the extinction coefficient (Fig. 1 (b)) along the a and b axes (3.2 µm-1) is larger than along the c-axis (2.9 $\mu$m$^-1$).The difference in efficiency could be explained by a small difference in the coupling strength to the electronic degrees of freedom. This is interesting, as the charge and orbital order develop within the ab-plane and one might intuitively expect that atomic motion within this plane would couple more strongly to the underlying electronic order.

\section*{Calculation of the phonon mode}
\begin{figure}
\centering
\includegraphics[scale=0.95]{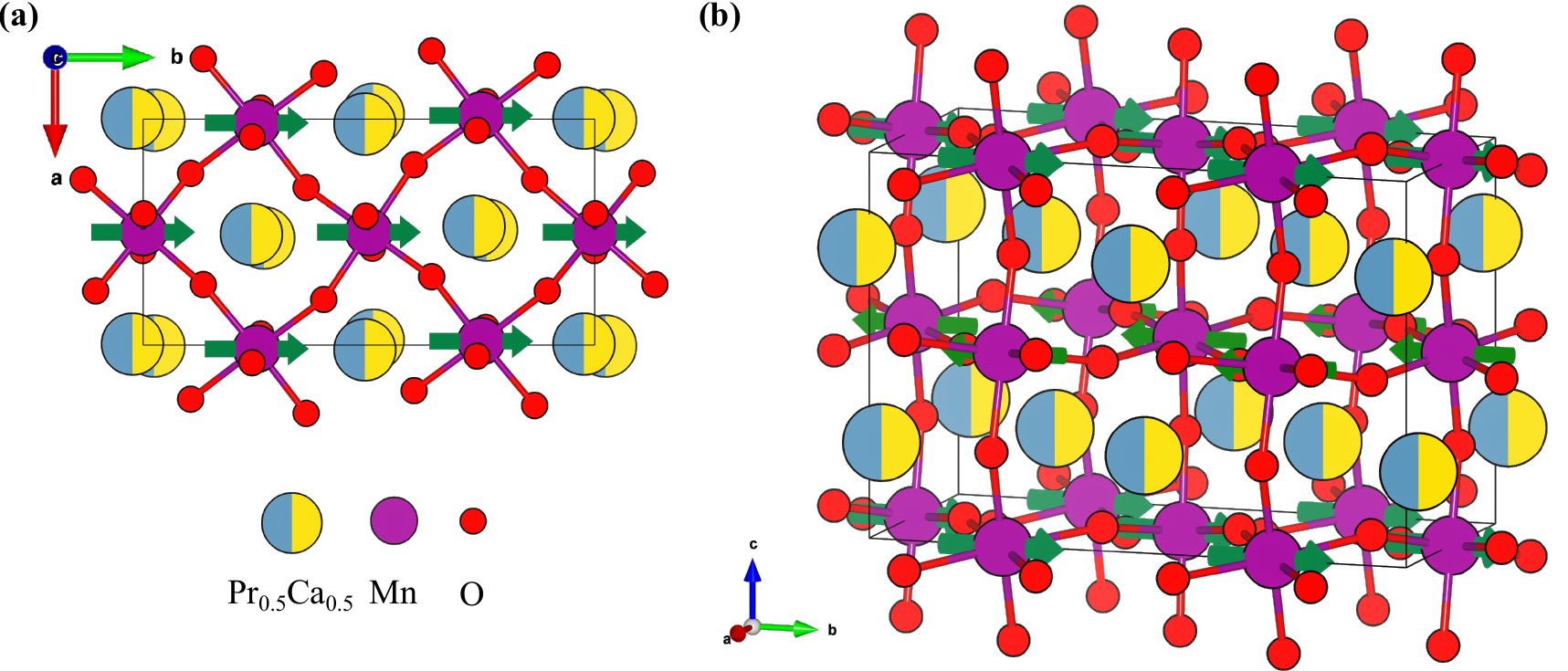}
\caption{Two different perspective on the PCMO unit cell. The green arrows on the Mn atoms represent the local spin arrangement for the A-type antiferromagnetic ordering.}
\label{S2}
\end{figure}

We treat the chemical 50/50 mixing of Pr and Ca in PCMO with a supercell approach, where either a Pr or Ca atoms occupy half of the A-sites positions within the 40 atoms unit cell. In total, we generated four unit cells with different distributions of Ca and Pr. Next, we compute for these four different configurations the total energies for different antiferromagnetic arrangements of the Mn spin moments. An A-type ordering (Fig. \ref{S2}) is found as the lowest energy magnetic ordering. The real magnetic unit cell is bigger than the supercell considered here and thus cannot be used. This however has only little influence to the result, a different magnetic ordering only slightly changing the size of the gap. For this magnetic structure we then relax the atomic positions within the unit cell. We list the structural data of one exemplary unit cell after the force minimization in Table S2. During the structural relaxation, the symmetry of the crystal was constrained to the $P2_1/m$ space group, which corresponds to the symmetry of the charge ordered (CO) state \cite{Goff2004}. Overall, our obtained structures exhibit a good agreement with the experimental data of Goff et al \cite{Goff2004}, also given in Table \ref{T2}.

\begin{table}[H]
\centering
\begin{tabular}{m{3cm} m{1.8cm} m{1.8cm} m{1.8cm} m{1.8cm} m{1.8cm} m{1.8cm} m{1.8cm}}
\hline \hline
Lattice constants & a (DFT) & a (EXP) & b (DFT) & b (EXP) & c (DFT) & c (EXP) \\ \hline
& 10.91 \r{A} & 10.87 \r{A} & 7.52 \r{A} & 7.49 \r{A} & 5.44 \r{A} & 5.43 \r{A} \\ \hline
Atomic positions & \\ \hline
Atom & Wykoff position & x (DFT) & x (EXP) & y (DFT) & y (EXP) & z (DFT) & z (EXP) \\ \hline
Pr/Ca &	e &	-0.012 &	-0.014 &	0.250 &	0.250 &	0.002 &	0.008 \\
Pr/Ca &	e &	-0.268 &	-0.266 &	0.250 &	0.250 &	0.485 &	0.490 \\
Pr/Ca &	e &	0.477 &	0.485 &	0.250 &	0.250 &	0.011 &	0.009 \\
Pr/Ca &	e &	0.232 &	0.238 &	0.250 &	0.250 &	0.499 &	0.491 \\
Mn1 &	f &	0.255 &	0.246 &	0.498 &	0.492 &	0.008 &	0.008 \\
Mn2a &	c &	0.000 &	0.000 &	0.000 &	0.000 &	0.500 &	0.500 \\
Mn2b &	d &	0.500 &	0.500 &	0.000 &	0.000 &	0.500 &	0.500 \\
O1 &	e &	0.001 &	0.008 &	0.250 &	0.250 &	-0.434 &	-0.436 \\
O2 &	e &	-0.239 &	-0.249 &	0.250 &	0.250 &	-0.078 &	-0.075 \\
O3 &	e &	-0.487 &	-0.492 &	0.250 &	0.250 &	-0.415 &	-0.422 \\
O4 &	e &	0.256 &	0.259 &	0.250 &	0.250 &	-0.074 &	-0.066 \\
O5 &	f &	0.355 &	0.360 &	0.460 &	0.461 &	0.299 &	0.293 \\
O6 &	f &	-0.110 &	-0.106 &	-0.464 &	-0.452 &	-0.224 &	-0.224 \\
O7 &	f &	-0.139 &	-0.144 &	0.464 &	0.469 &	0.263 &	0.262 \\
O8 &	f &	0.392 &	0.386 &	-0.455 &	-0.468 &	-0.212 &	-0.208 \\ 
\hline \hline
\end{tabular}
\caption{Comparison of experimental (EXP) \cite{Goff2004} and calculated in this work (DFT) structural parameters of PCMO.}
\label{T2}
\end{table}

Next, we compute the phonon frequencies at the zone center (q=(0,0,0)) for all four structures by calculating the force constant matrix using a finite difference approach as implemented in the Phonopy software package \cite{Togo2015}. The 40 atoms unit cell exhibit 117 non-translational modes which are spanned by the irreducible representations of the 2/m point group. There are explicitly 31 A$_g$ modes, 28 A$_u$ modes, 35 B$_u$ modes and 23 B$_g$ modes. We note that the A$_u$ and B$_u$ modes exhibit a polarization along the \textit{c} and \textit{a-b} axis, respectively.

\begin{figure}
\centering
\includegraphics[scale=0.72]{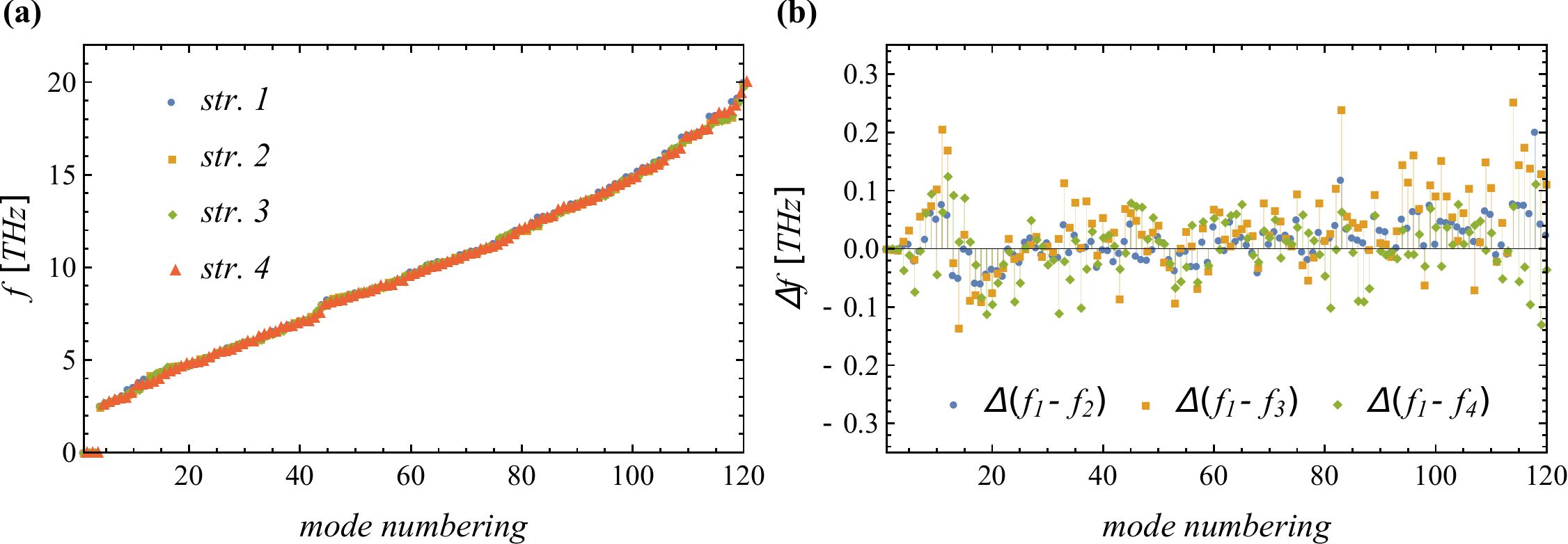}
\caption{Comparison of the phonon mode frequencies at q=(0,0,0) for four different distributions of Pr and Ca atoms within the PCMO structure. (a) Frequency as a function of mode numbering for all four structures and (b) the difference of phonon mode frequencies with respect to structure 1, which is given in Table \ref{T2}.}
\label{S3}
\end{figure}

We show the computed phonon frequencies for the four distinct distributions of Pr and Ca atoms in Fig. \ref{S3} (a). All four structures exhibit very similar phonon frequencies. To quantify this observation, we computed further how the frequencies of the phonon modes changes between the four structures and show these differences in Fig. \ref{S3} (b). Each specific mode frequency does not vary more than 0.3 THz between the different structural configurations. Consequently, due to the insensitivity of the phononic system of PCMO on the Pr and Ca distribution, we concentrate in the following our discussion on one configuration with its structure given in Table \ref{T2}.

\begin{figure}
\centering
\includegraphics[scale=0.65]{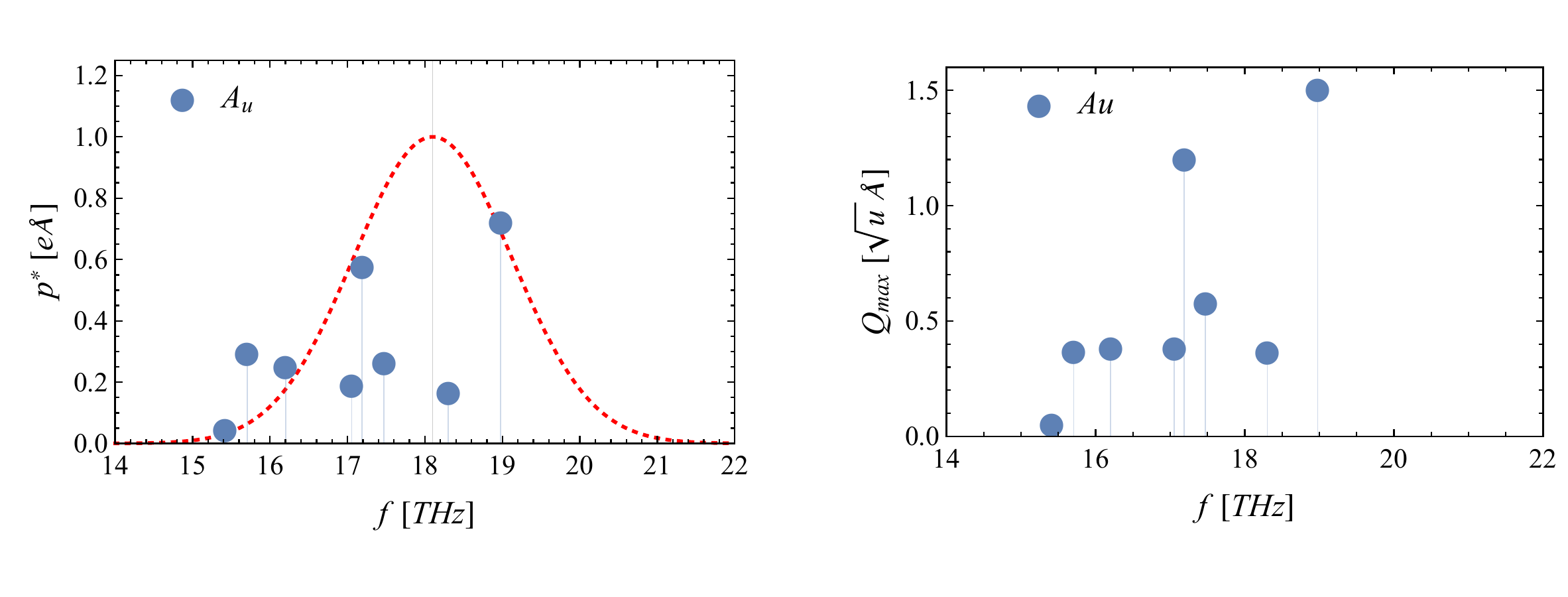}
\caption{(a) A$_u$ mode polarity $p^*$ in the range of the pulse frequency. The dashed red curve corresponds to the frequency width of the experimental pulse. (b) Resulting maximal mode displacement of polar A$_u$ modes.}
\label{S4}
\end{figure}

In the next step, we determine the mode displacement due to the excitation by a THz pulse. We thereby follow the approach of Jurascheck et al. \cite{Juraschek2016} and solve the dynamical equation for a harmonic oscillator driven by a pulsed excitation. The resulting displacement of the mode depends on both the mode specific dipole moment \textit{p*} and the matching of pulse bandwidth and mode frequencies. We show in Fig. \ref{S3} (a),(b) the mode polarities and the resulting displacements of polar modes due to the excitation with the THz pump pulse provided during the experiment. The data shown are restricted to the A$_u$ modes with the polarization along the c-axis corresponding to the experiment. Two modes exhibit a significant dipole moment (Fig. \ref{S4} (a)) and are found to be significantly excited by the THz pulse (Fig. \ref{S4} (b)). The main anharmonic displacements and changes in the electronic structure will be driven by the largest polar modes displacements. Consequently, we focus in the following on these two modes with frequencies 17.2 THz and 18.9 THz.

\section*{Anharmonic phonon coupling}
\begin{figure}
\centering
\includegraphics[scale=0.88]{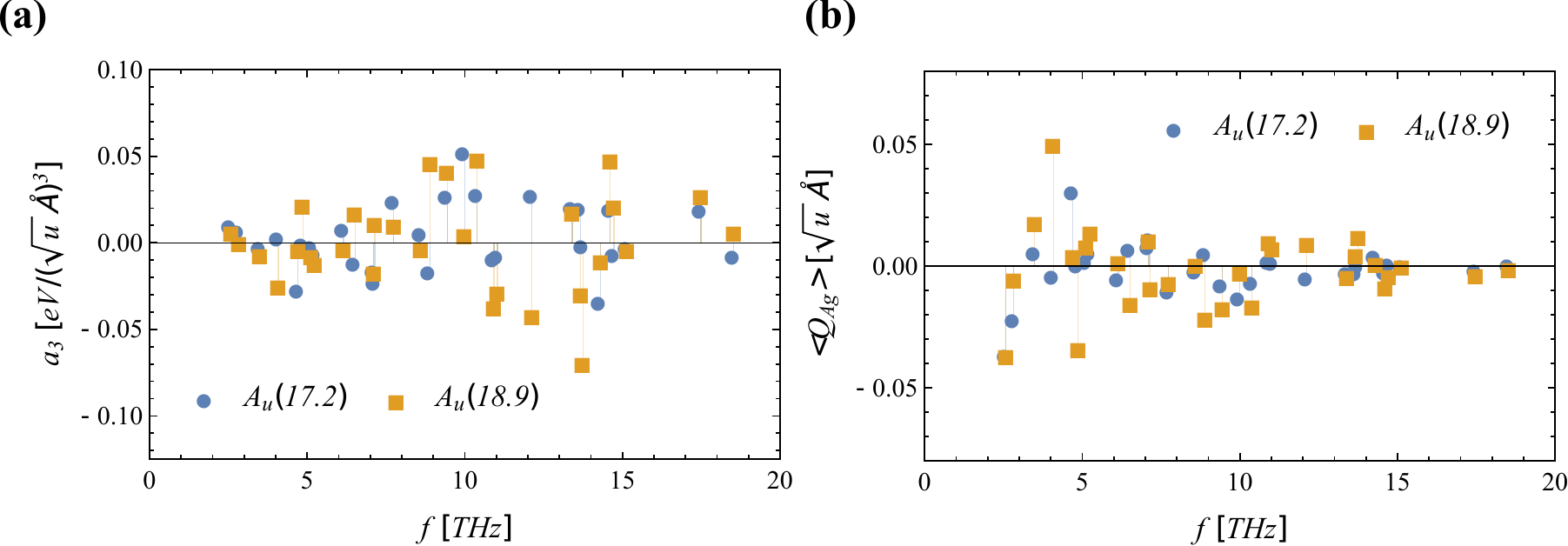}
\caption{(a) Cubic anharmonic coupling constants of the A$_u$(17.2) and A$_u$(19.9) polar modes with all other phonon modes of A$_g$ symmetry.  (b) Maximal amplitude of the A$_g$ mode induced by the anharmonic coupling and displacement of the polar mode.}
\label{S5}
\end{figure}

We also explore the anharmonic coupling of the two excited polar modes to other modes, comparable to the analysis by Subedi et al \cite{Subedi2014}. We restrict our analysis to the cubic coupling of potential given in the main text. Similar to Refs. \cite{Fechner2016, Juraschek2016}, we compute $a_3$ by performing frozen phonon calculations and fitting the resulting energy landscape to the potential given in the main text. We note that the cubic coupling is only allowed between the polar mode and modes of A$_g$ symmetry. Consequently, we performed 62 separate calculations for each pair of polar and A$_g$ mode.
Figure \ref{S5} (a) shows an overview of the obtained $a_3$ coupling constants over the entire frequency range. We note that the found cubic coupling constants are small compared to other materials as given in Refs \cite{Subedi2014, Fechner2016, Juraschek2016}. Cubic anharmonicities between polar and A$_g$ modes cause transient structural modulations by the A$_g$ mode which are driven by the polar mode displacement \cite{Fechner2016}. The size of these modulations is given by
\begin{equation}
\langle Q_{\text{A$_g$}} \rangle = \frac{a_3 Q_{\text{IR}}^2}{\omega_{\text{A$_g$}}^2} \text{.}
\label{E1}
\end{equation}
Thus beside the coupling constants, the excitation strength of the polar mode and the mode frequency of the Ag mode also influence the displacement of the A$_g$ mode. To quantify the distortion created by the cubic anharmonicities we evaluate Eqn. \ref{E1} for our computed coupling constants in combination with the amplitudes of the polar mode created by the largest fluences. Figure \ref{S5} (b) shows the resulting displacements of Ag modes created by both polar modes. Overall, we find that cubic anharmonicities only create negligible to small displacements of the A$_g$ modes. Mode amplitudes of 0.05 $\sqrt{\mu}$/\r{A} correspond to average atomic displacements of 0.02 pm, which are too small to exhibit a sizeable effect.

\begin{figure}
\centering
\includegraphics[scale=0.88]{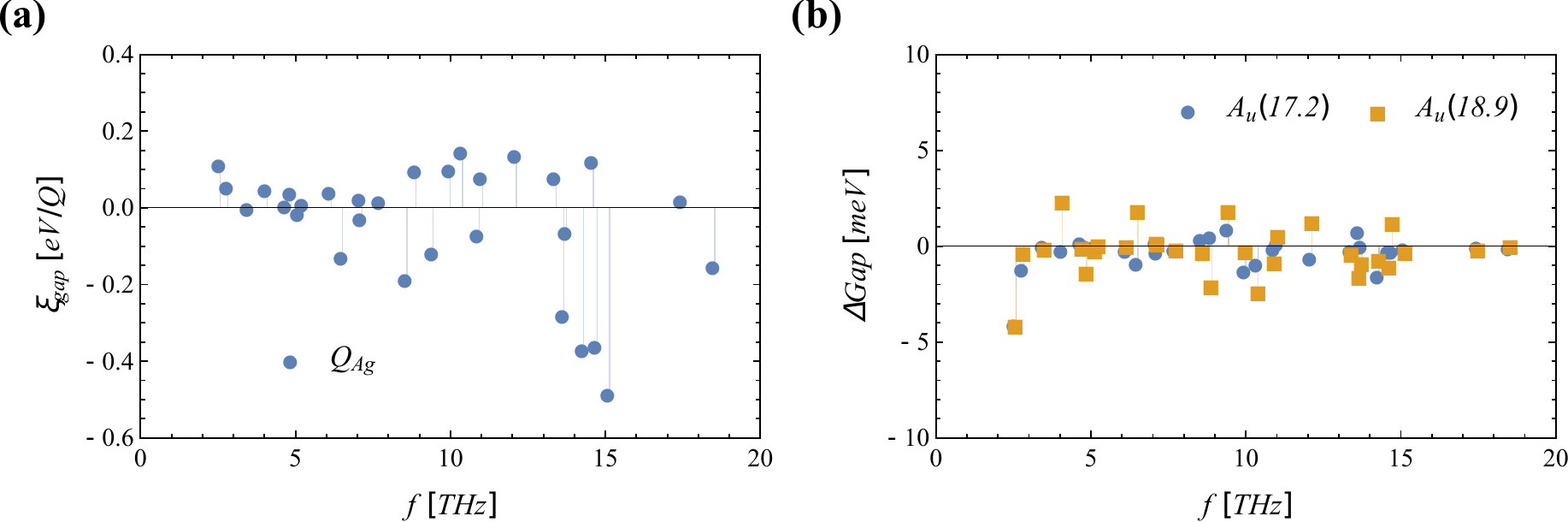}
\caption{ Effects of the induced Raman displacement on the gap. (a) Linear coefficient of the mode amplitude dependent gap for all calculated Raman modes. (b) Gap change as a function of the induced Raman displacement by both excited modes.}
\label{S6}
\end{figure}

\begin{figure}
\centering
\includegraphics[scale=1.3]{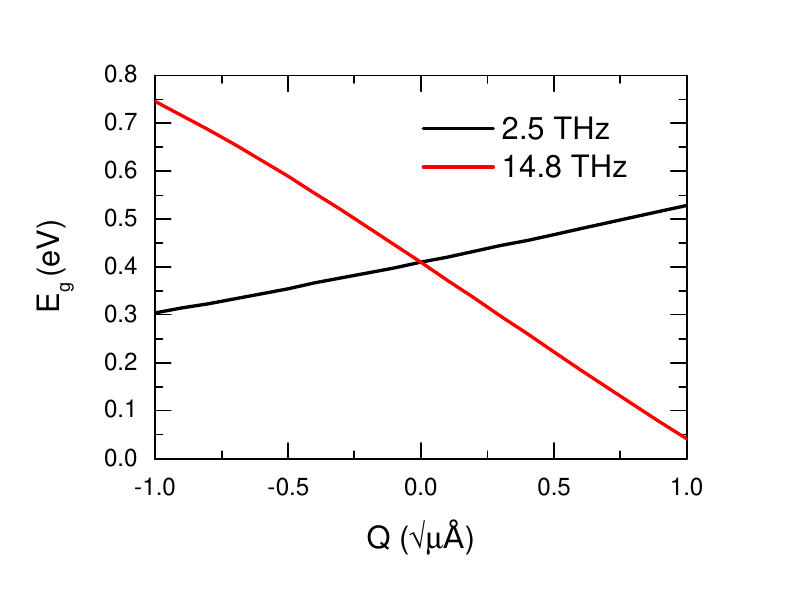}
\caption{Gap as a function of two Raman modes amplitude. A mostly linear dependence is observed, as expected from the mode symmetries.}
\label{S7}
\end{figure}

Although the induced distortions are very small, we checked the behavior of the gap as a function of the Raman modes amplitudes and their corresponding displacements. These results are summarized in Fig. \ref{S6}. Some modes indeed couple strongly to the gap (Fig. \ref{S6} (a)); however these are high frequency modes whose displacements are tiny, resulting in a negligible change of the gap energy (Fig. \ref{S6} (b)). In Figure \ref{S7}, we show the gap as a function of the mode amplitude for two Raman phonons: a mode at 2.4 THz, which is displaced the most and a mode at 14.8 THz, which has a large effect on the gap. The gap depends linearly on the mode amplitude and the direction of the displacement leads to a reduction of the gap, as predicted by Subedi et al. in Ref. \cite{Subedi2014}. However, the overall contribution of the Raman modes is estimated to -15 meV, clearly too small to induce the transition to the metallic state.

\bibliography{./ref_midIR}

\end{document}